\documentclass[conference]{IEEEtran}

\usepackage[T1]{fontenc}
\usepackage{cite}
\usepackage{amsmath,amssymb,amsfonts}
\usepackage{graphicx}
\usepackage{booktabs}
\usepackage{multirow}
\usepackage{array}
\usepackage{url}
\usepackage{balance}
\usepackage{algorithm}
\usepackage{algpseudocode}
\usepackage{siunitx}
\usepackage{xcolor}
\usepackage{tikz}
\usepackage{eso-pic}
\usetikzlibrary{arrows.meta,positioning,fit,shapes.geometric,calc}
\usepackage{pgfplots}
\pgfplotsset{compat=1.18}
\usepackage[hidelinks]{hyperref}

\title{PRISM: A Post-Prediction Reliability Framework for
Evidence-Guided Correction and Selective Abstention}

\author{
Omm~Swarup\textsuperscript{1,2},
Shrishail~M.~Hiremath\textsuperscript{2}\\
\textsuperscript{1}Indian Institute of Technology Madras,\\
Chennai, Tamil Nadu, India\\
\textsuperscript{2}Department of Electronics and Communication Engineering,\\
National Institute of Technology Rourkela,\\
Rourkela, Odisha, India\\
Email: \{24f2000745@ds.study.iitm.ac.in, 125ei0038@nitrkl.ac.in, hiremaths@nitrkl.ac.in\}
}

\begin{document}
\maketitle


\AddToShipoutPictureFG{%
    \AtPageLowerLeft{%
        \put(18,390){%
            \rotatebox{90}{%
                \textcolor{gray!35}{%
                    \sffamily\bfseries\fontsize{11}{13}\selectfont
                    PRISM PREPRINT 
                }%
            }%
        }%
    }%
}

\begin{abstract}
A model may produce a prediction with high confidence and still get the outcome wrong. This is a problem in sequential applications, where the system evolves over time and the data observed during deployment may differ from what the model observed during training. To address this issue we propose PRISM (Prediction Reliability and Intelligent Self-Correction Mechanism), a framework that evaluates the reliability of a prediction after it has been generated. It first estimates the possibility of an error and then examines the predictions identified as potentially unreliable. For such cases, PRISM analyses the available signals, searches for similar situations in historical data and checks whether an alternative prediction is supported by the available evidence. Based on the result of this process, it can retain the original prediction, replace it with a verified correction or abstain when the evidence is not enough to make a reliable decision.

We study PRISM by a match's 2nd innings prediction in the Indian Premier League (IPL). Ball by ball records are converted into over level observations to study how the state of a match changes over an innings. To preserve the temporal order of the data, earlier IPL seasons are used for training, IPL 2024 is reserved for validation, and IPL 2025–2026 are used for out of time testing. The documented dataset consists of 8,665 over level observations from 475 matches, with the reported test partition containing 2,568 observations from 142 matches. In one experimental record, Logistic Regression, Random Forest and HistGradientBoosting achieve accuracies of 65.2\%, 68.4\% and 70.2\%, respectively. The corresponding PRISM results show a selective accuracy of 82.4\% at 88.5\% coverage with a Brier score of 0.126 and an expected calibration error (ECE) of 0.019.
\end{abstract}


\section{Introduction}

Accuracy and log loss are commonly used to measure the performance of
machine learning models. While these metrics help compare different
approaches, they do not fully address an important question i.e how should
a system respond when it comes across a prediction that may not be
reliable? This question becomes important in sequential
applications.

A model's confidence can provide useful information about it's
predictions but prediction with a high confidence is not necessarily a correct one. Calibration methods try to make predicted probabilities more consistent with the frequencies observed. Temperature
scaling, for example, is a commonly used post-hoc calibration method
\cite{guo2017}. Selective classification uses a different approach by
allowing a model to withhold its prediction when the estimated risk is
too high \cite{geifman2017,shafer2008}. Measures like the Brier score
also help assess the quality of predicted probabilities rather than
considering accuracy alone \cite{brier1950,gnr2007}. In addition,
out of distribution detection and ensemble based uncertainty estimation
have been used until now to identify unfamiliar inputs and quantify
uncertainty in model outputs \cite{hendrycks2017,lakshminarayanan2017}.
Although these methods provide ways to identify potentially unreliable
predictions, determining whether an incorrect prediction can be
corrected using available evidence remains a separate challenge.

PRISM (Prediction Reliability and Intelligent Self-Correction
Mechanism) is proposed to address this problem. It works after a base
model has generated its prediction and examines whether that output
should be trusted. The framework estimates the risk of error, investigates
the signals associated with a potentially unreliable prediction and
finds relevant examples from historical data. Before an alternative
is accepted, it is passed through a verification stage to determine
whether the available evidence supports the proposed change. The system
don't attempt to correct every prediction it flags. When the evidence
is not sufficient, it can abstain rather than introduce a correction that
may be less reliable than the original output.

The main contributions of this study are:
\begin{itemize}
    \item We propose PRISM, a framework that performs
    reliability assessment and intervention after the initial
    prediction has been generated.

    \item We introduce an error diagnosis and historical
    evidence retrieval process to identify possible alternatives
    for predictions considered unreliable.

    \item We use a verification phase that determines
    whether a proposed correction should be accepted with the
    final decision allowing the system to retain, correct or
    abstain from a prediction.

    \item We evaluate this framework on 2nd innings IPL
    prediction using chronological match level splits, prediction coverage and correction related metrics.
\end{itemize}

\section{Related Work}

\subsection{Calibration and Predictive Uncertainty}
Calibration examines whether predicted probabilities agree with actual
outcomes. Guo \emph{et al.} introduced temperature scaling as a simple
way to improve neural network calibration \cite{guo2017}, while
Lakshminarayanan \emph{et al.} explored deep ensembles for estimating
predictive uncertainty \cite{lakshminarayanan2017}. PRISM uses these
ideas as reliability signals since confidence alone cannot guarantee
a correct prediction.

\subsection{Selective Prediction}
Selective prediction enables a model to abstain when the estimated risk
is high, leading to a trade off between coverage and prediction accuracy.
Geifman and El-Yaniv explored this approach for deep neural networks
\cite{geifman2017}. PRISM extends this idea by retrieving historical
examples and verifying a possible correction before accepting it.

\subsection{Out-of-Distribution Detection}
Out of distribution detection identifies inputs that differ from the
training data. Hendrycks and Gimpel proposed maximum softmax
probability as a simple detection baseline \cite{hendrycks2017}.
Although such methods help flag unfamiliar inputs, they do not
determine whether a prediction should be corrected or rejected.
PRISM solves this through evidence retrieval and verification.

\section{Problem Formulation}
Let $\mathcal{D}=\{(x_i,y_i)\}_{i=1}^{N}$ denote a dataset of sequential
prediction states, where $x_i\in\mathcal{X}$ contains the information available
at a particular instance and $y_i\in\{0,1\}$ denotes the eventual outcome.
A base model $f_{\theta}$ produces
\begin{equation}
    \hat{p}_i=f_{\theta}(x_i), \qquad
    \hat{y}_i=\mathbb{I}[\hat{p}_i\geq 0.5].
\end{equation}
PRISM receives $x_i$, $\hat{p}_i$, and inference time signals. It
returns a decision
\begin{equation}
    d_i\in\{\mathrm{ACCEPT},\mathrm{CORRECT},\mathrm{ABSTAIN}\}.
\end{equation}
The final prediction is defined only when the system accepts the original
prediction or an alternative:
\begin{equation}
\tilde{y}_i =
\begin{cases}
\hat{y}_i, & d_i=\mathrm{ACCEPT},\\
\hat{y}_{c,i}, & d_i=\mathrm{CORRECT},\\
\bot, & d_i=\mathrm{ABSTAIN},
\end{cases}
\end{equation}
where $\bot$ denotes abstention and $\hat{y}_{c,i}$ is a candidate correction.
The evaluation reports both the quality of accepted predictions
and the fraction of inputs for which a prediction is returned.

\section{PRISM Framework}
\subsection{Overview}
\begin{figure*}[t]
\centering
\begin{tikzpicture}[
  font=\small,
  block/.style={draw,rounded corners=2pt,thick,align=center,
    minimum height=10mm,text width=23mm,inner sep=3pt,fill=blue!5},
  stage/.style={draw,rounded corners=2pt,thick,align=center,
    minimum height=11mm,text width=26mm,inner sep=3pt,fill=cyan!8},
  decision/.style={draw,rounded corners=2pt,thick,align=center,
    minimum height=10mm,text width=25mm,inner sep=3pt,fill=orange!12},
  output/.style={draw,rounded corners=2pt,thick,align=center,
    minimum height=9mm,text width=24mm,inner sep=3pt,fill=green!8},
  arr/.style={-{Latex[length=2.2mm]},thick},
  fb/.style={-{Latex[length=2mm]},thick,dashed}
]
\node[block]   (input) at (0,0) {Match state\\$x_t$};
\node[block]   (base)  at (3.2,0) {Base predictor\\$f_\theta$};
\node[block]   (pred)  at (6.4,0) {Original output\\$\hat p_t,\hat y_t$};
\node[stage]   (detect)at (9.6,0) {Detect\\Error-risk score $r_t$};
\node[stage]   (diag)  at (12.8,0) {Diagnose\\Failure signals};
\draw[arr] (input)--(base);
\draw[arr] (base)--(pred);
\draw[arr] (pred)--(detect);
\draw[arr] (detect)--(diag);
\node[stage] (retrieve) at (12.8,-2.0) {Correct\\Retrieve historical states\\and propose $\hat y_{c,t}$};
\node[stage] (verify) at (9.6,-2.0) {Verify\\Check evidence and\\candidate validity};
\node[decision] (controller) at (6.4,-2.0) {Decision controller\\Risk and verification rules};
\draw[arr] (diag)--(retrieve);
\draw[arr] (retrieve)--(verify);
\draw[arr] (verify)--(controller);
\node[output] (accept) at (2.4,-4.0) {ACCEPT\\Keep $\hat y_t$};
\node[output] (correct) at (6.4,-4.0) {CORRECT\\Use $\hat y_{c,t}$};
\node[output] (abstain) at (10.4,-4.0) {ABSTAIN\\Return $\bot$};
\draw[arr] (controller.south) -- ++(0,-5mm) -| (accept.north);
\draw[arr] (controller.south) -- (correct.north);
\draw[arr] (controller.south) -- ++(0,-5mm) -| (abstain.north);
\draw[fb] (retrieve.west) -- ++(-8mm,0) |- node[pos=0.25,left,font=\scriptsize,align=right] {retrieved evidence} (diag.south);
\end{tikzpicture}
\caption{PRISM architecture. The upper row generates and diagnoses the original
prediction. The lower row performs historical retrieval and independent
verification before the controller selects acceptance, correction, or abstention.}
\label{fig:arch}
\end{figure*}
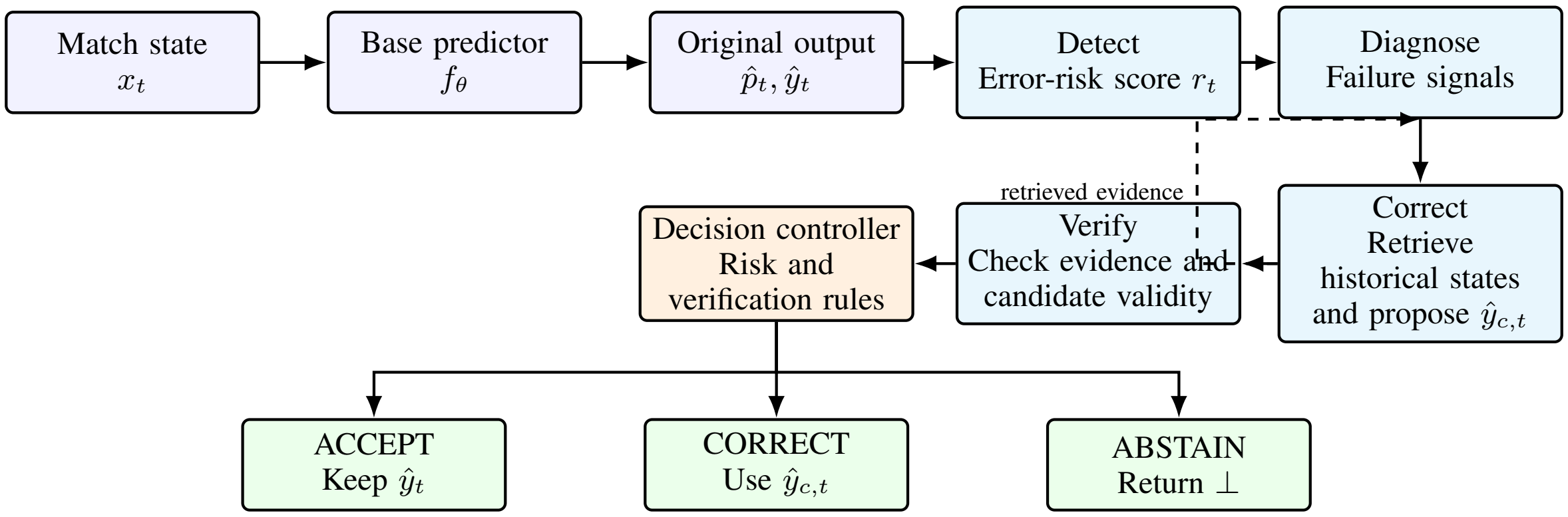

PRISM is divided into 5 stages: Detect, Diagnose, Correct, Verify and Decide. The base predictor remains a separate component. Fig.~\ref{fig:arch} shows the workflow.

\subsection{Detection}

The first stage of PRISM examines the prediction made by the base
model and estimates the probability of it being incorrect. For this
purpose, the detector uses a set of signals $\phi_i$, which 
includes prediction confidence, probability margin, uncertainty
estimates, similarity scores and model disagreement. The estimated
error risk is given by
\begin{equation}
    r_i=g_{\psi}(\phi_i), \qquad r_i\in[0,1],
\end{equation}
where a higher value of $r_i$ indicates a greater probability of error.
A threshold $\tau_r$ is then used to decide whether the prediction
requires further correction:
\begin{equation}
    q_i=\mathbb{I}[r_i>\tau_r].
\end{equation}
To prevent the detector from learning overly optimistic error
estimates, its training targets are generated using out of fold or
held out predictions rather than predictions obtained directly from
the training samples.

\subsection{Diagnosis}

Once a prediction is flagged, PRISM examines the corresponding input
to understand the context in which the prediction was made. This stage
forms a diagnostic representation using the input state, the original
prediction and its estimated risk:
\begin{equation}
    z_i=h_{\omega}(x_i,\hat{p}_i,r_i).
\end{equation}
For the IPL application, this representation describes the current
match situation and provides a basis for comparing it with previously
observed matches.

\subsection{Evidence Guided Correction}

PRISM uses previously observed examples as a source of evidence when
considering a possible correction. Its reference memory is defined as
\begin{equation}
    \mathcal{M}=\{(u_j,y_j)\}_{j=1}^{M},
\end{equation}
where $u_j$ represents a historical state and $y_j$ denotes its
eventual outcome. For a flagged prediction, the retrieval module
searches this memory for the $K$ states most similar to the current representation:
\begin{equation}
    \mathcal{N}_{K}(z_i)=
    \operatorname{KNN}(z_i,\mathcal{M},K).
\end{equation}
The experiments use $K=50$ neighbours. These retrieved examples are
then used to propose an alternative prediction:
\begin{equation}
    \hat{y}_{c,i}=C_{\omega}
    \left(x_i,\hat{y}_i,\mathcal{N}_{K}(z_i)\right).
\end{equation}
The retrieved evidence is used to support the correction rather than
replace the base model directly. To maintain a valid evaluation,
the reference memory contains only examples from the training period.

\subsection{Verification}

Before a proposed correction is accepted, PRISM checks whether it is
consistent with the available evidence. This is important because
similar historical examples do not necessarily mean that the
alternative prediction is correct. The verification stage evaluates
the original prediction, the proposed alternative and the retrieved
neighbours:
\begin{equation}
    v_i=V_{\eta}
    \left(x_i,\hat{y}_i,\hat{y}_{c,i},\mathcal{N}_{K}(z_i)\right).
\end{equation}
The resulting score $v_i$ indicates the support level of the
candidate correction. A correction is considered only when it exceeds
the required verification threshold. The verification rule and its
threshold are selected using validation data and kept fixed during
testing.

\subsection{Decision and Abstention}

The final stage combines the estimated error risk with the verification
result to determine what should happen to the original prediction.
PRISM supports three possible outcomes: accepting the original
prediction, replacing it with a verified correction or abstaining
when neither decision is sufficiently supported.
\begin{equation}
d_i =
\begin{cases}
\mathrm{ACCEPT}, & r_i\leq\tau_r,\\
\mathrm{CORRECT}, & r_i>\tau_r \land v_i\geq\tau_v,\\
\mathrm{ABSTAIN}, & \text{otherwise}.
\end{cases}
\end{equation}
In this way, PRISM does not force a correction whenever the detector
identifies a potentially unreliable prediction.

\begin{algorithm}[t]
\caption{PRISM inference procedure}
\label{alg:prism}
\begin{algorithmic}[1]
\Require State $x_i$, base model $f_\theta$, memory $\mathcal{M}$
\Ensure Decision $d_i$ and optional final prediction $\tilde y_i$
\State $\hat p_i\gets f_\theta(x_i)$
\State $\hat y_i\gets\mathbb{I}[\hat p_i\geq 0.5]$
\State $\phi_i\gets\operatorname{ExtractSignals}(x_i,\hat p_i)$
\State $r_i\gets g_\psi(\phi_i)$
\If{$r_i\leq\tau_r$}
    \State \Return $(\mathrm{ACCEPT},\hat y_i)$
\EndIf
\State $z_i\gets h_\omega(x_i,\hat p_i,r_i)$
\State $\mathcal{N}_K\gets\operatorname{Retrieve}(z_i,\mathcal{M},K)$
\State $\hat y_{c,i}\gets C_\omega(x_i,\hat y_i,\mathcal{N}_K)$
\State $v_i\gets V_\eta(x_i,\hat y_i,\hat y_{c,i},\mathcal{N}_K)$
\If{$v_i\geq\tau_v$}
    \State \Return $(\mathrm{CORRECT},\hat y_{c,i})$
\Else
    \State \Return $(\mathrm{ABSTAIN},\bot)$
\EndIf
\end{algorithmic}
\end{algorithm}

\section{Experimental Setup}

\subsection{Dataset and Prediction}

The experiments use ball by ball IPL data obtained from Cricsheet
\cite{cricsheet}. The objective is to predict the outcome of a match
during its 2nd innings. For this, the delivery level records
are converted into over level data, each representing the match
situation at a particular stage of the chase. Only information available
up to that point is used to construct each data.

\subsection{Chronological Partitioning}

To reflect how the model would be used in practice matches are divided
chronologically rather than randomly:
\begin{itemize}
    \item Training: IPL seasons 2020/2021--2023.
    \item Validation: IPL 2024.
    \item Testing: IPL 2025--2026.
\end{itemize}

\begin{figure}[t]
\centering
\begin{tikzpicture}[font=\scriptsize, x=0.95cm, y=0.9cm,
  period/.style={draw,rounded corners=2pt,minimum height=8mm,align=center,thick},
  flow/.style={-{Latex[length=2mm]},thick}]
\node[period,fill=blue!8,text width=22mm] (tr) at (0,0) {Training\\2020/21--2023};
\node[period,fill=orange!10,text width=18mm] (va) at (3.2,0) {Validation\\2024};
\node[period,fill=green!10,text width=22mm] (te) at (6.1,0) {Out-of-time test\\2025--2026};
\draw[flow] (tr)--(va);
\draw[flow] (va)--(te);
\node[align=center] at (3.1,-0.75) {Chronological, match-level split\\All states from a match stay in one partition};
\end{tikzpicture}
\caption{Documented chronological evaluation protocol for the IPL case study.}
\label{fig:split}
\end{figure}
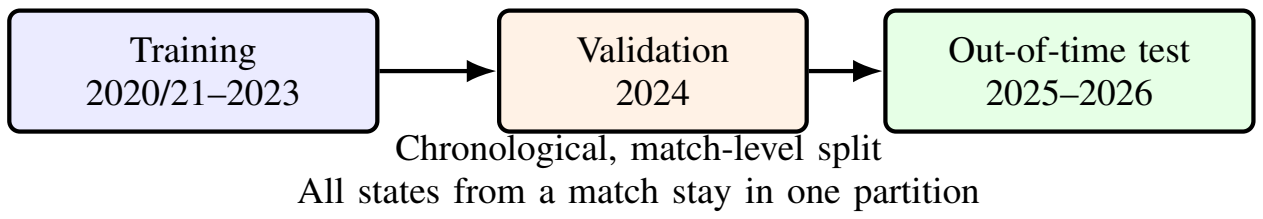

The partitioning strategy is shown in Fig.~\ref{fig:split}. All
observations belonging to a particular match remain in the same
partition, preventing information from a match from appearing across
training and testing.

\subsection{Baseline Models}

We compare PRISM against three commonly used classifiers:
Logistic Regression (LR), Random Forest (RF) and HistGradientBoosting
(HGB). These models provide three different approaches to classification:
a linear probabilistic model, a bagging based tree ensemble and a
gradient boosted tree model, respectively
\cite{breiman2001,friedman2001,hastie2009}. Their preprocessing
procedures, feature sets, hyperparameters and random seeds are
reported to make the experiments reproducible.

\subsection{Machine Learning Methodology}
\label{subsec:mlmethod}

The prediction task is treated as binary classification at each
stage of the 2nd innings. Let $x_{m,o}\in\mathbb{R}^{d}$ represent
the feature vector for match $m$ after over $o$ and let
$y_m\in\{0,1\}$ indicate whether the chasing team eventually wins.
Each feature vector is constructed using information available up to
the corresponding over. Future deliveries, final scores, match
identifiers and post match statistics are not included to avoid target
leakage.

\subsubsection{Base Classifiers}

Logistic Regression is used as a linear probabilistic baseline. It
estimates the probability that the chasing team wins as
\begin{equation}
 p_{\mathrm{LR}}(y=1\mid x)
 =\sigma(\beta_0+\beta^\top x),\qquad
 \sigma(a)=\frac{1}{1+e^{-a}},
\end{equation}
where the parameters are learned by minimizing binary cross entropy,
with $L_2$ regularization.

Random Forest combines multiple decision trees trained on bootstrap
samples and randomized feature subsets. Its final probability estimate
is obtained by averaging the predictions of the individual trees:
\begin{equation}
 p_{\mathrm{RF}}(y=1\mid x)
 =\frac{1}{T}\sum_{t=1}^{T}p_t(y=1\mid x),
\end{equation}
where $T$ denotes the number of trees \cite{breiman2001}.

HistGradientBoosting builds trees sequentially with each new tree
helping to reduce the errors made by the existing ensemble
\cite{friedman2001}. For binary classification, it uses logistic loss
and converts the final model output into a probability.

All three classifiers are trained using the earlier seasons, selected
on the validation data and evaluated on the later seasons. Their
hyperparameters, preprocessing settings, random seeds and calibration
configurations must be recorded for reproducibility.

\subsubsection{Error Detector Learning}

PRISM uses a second classifier to estimate whether the base model is
likely to make an error. For each training observation, the error
indicator is defined as
\begin{equation}
 e_i=\mathbb{I}[\hat y_i\ne y_i].
\end{equation}
The detector uses inference time signals $\phi_i$,
predicted probability, probability margin, uncertainty estimates and
similarity to historical states. It's output represents the estimated
probability of an incorrect prediction:
\begin{equation}
 r_i=P(e_i=1\mid\phi_i).
\end{equation}

A higher value of $r_i$ indicates that the prediction is highly likely to be incorrect. The risk threshold
$\tau_r$ is selected using validation data according to the chosen
risk--coverage objective and is not adjusted using test outcomes.

\subsubsection{Historical Retrieval and Candidate Generation}

When a prediction is flagged, PRISM searches for similar match situations. The current state is first represented
by
\begin{equation}
 z_i=h_\omega(x_i,\hat p_i,r_i).
\end{equation}
This representation is compared with historical data $u_j$.
A normalized Euclidean distance is used to measure their
similarity:
\begin{equation}
 d(z_i,u_j)=
 \sqrt{\sum_{k=1}^{d}
 \left(\frac{z_{ik}-u_{jk}}{s_k}\right)^2},
\end{equation}
where $s_k$ is a feature wise scale calculated from training data.
The $K$ closest observations are retrieved with $K=50$ in the
recorded experiments.

The retrieved outcomes are then used to estimate a candidate
probability through distance weighted voting:
\begin{equation}
\begin{aligned}
 p_{\mathrm{KNN}}(y=1\mid z_i)
 &=\frac{\sum_{j\in\mathcal{N}_K(z_i)}w_{ij}y_j}
 {\sum_{j\in\mathcal{N}_K(z_i)}w_{ij}},\\
 w_{ij}&=\frac{1}{d(z_i,u_j)+\epsilon}.
\end{aligned}
\end{equation}
Here, $\epsilon>0$ is to avoid division by 0. The candidate label is
obtained as
\begin{equation}
 \hat y_{c,i}=
 \mathbb{I}[p_{\mathrm{KNN}}(y=1\mid z_i)\geq 0.5].
\end{equation}

This formulation describes a distance weighted KNN candidate
generator. The retrieval memory is restricted
to training period observations so that test outcomes cannot
influence the correction process \cite{cover1967}.

\subsubsection{Verification and Calibration}

A candidate generated from historical neighbours is not accepted
automatically. PRISM uses a verification stage to examine whether
the proposed alternative has sufficient support in the retrieved
evidence. This assessment considers the estimated risk, the
strength of the neighbour vote, agreement between retrieved labels,
distance statistics and consistency with the original prediction.
The resulting verification score is denoted by $v_i$.

A threshold $\tau_v$ determines whether the candidate can be
accepted as a correction. This threshold is selected using validation
data and remains fixed during testing. Established approaches such as temperature scaling and
Platt scaling can be fitted using held out validation predictions
\cite{guo2017,platt1999,niculescu2005}.

\subsection{Metrics}
Accuracy is computed as
\begin{equation}
\mathrm{Acc}=\frac{1}{N}\sum_{i=1}^{N}
\mathbb{I}[\hat y_i=y_i].
\end{equation}
For a set of accepted predictions $\mathcal{A}$, selective accuracy is
\begin{equation}
\mathrm{Acc}_{\mathrm{sel}}=
\frac{\sum_{i\in\mathcal{A}}\mathbb{I}[\tilde y_i=y_i]}
{|\mathcal{A}|},
\end{equation}
and coverage is
\begin{equation}
\mathrm{Cov}=\frac{|\mathcal{A}|}{N}.
\end{equation}
For binary probabilistic predictions, the Brier score is
\begin{equation}
\mathrm{BS}=\frac{1}{N}\sum_{i=1}^{N}(\hat p_i-y_i)^2.
\end{equation}
Expected calibration error is calculated using $B$ confidence bins, following common calibration practice \cite{guo2017,niculescu2005}:
\begin{equation}
\mathrm{ECE}=\sum_{b=1}^{B}\frac{|S_b|}{N}
\left|\operatorname{acc}(S_b)-\operatorname{conf}(S_b)\right|.
\end{equation}
Correction specific metrics should also be reported:
\begin{align}
\mathrm{EDR}&=\frac{\text{base errors detected}}{\text{all base errors}},\\
\mathrm{ERR}&=\frac{\text{base errors correctly recovered}}{\text{all base errors}},\\
\mathrm{HCR}&=\frac{\text{correct base predictions made incorrect}}
{\text{originally correct base predictions}}.
\end{align}

\section{Results}
\subsection{Baseline Comparison}

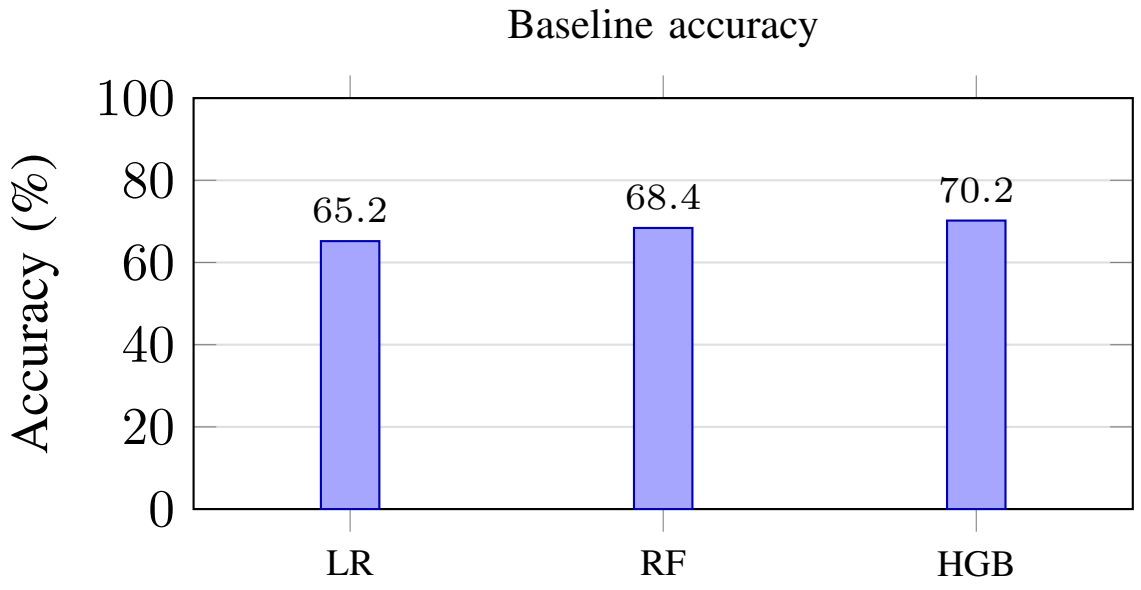
\begin{figure}[t]
\centering
\begin{tikzpicture}
\begin{axis}[
  ybar, bar width=11pt, width=0.88\columnwidth, height=4.3cm,
  ymin=0, ymax=100, ylabel={Accuracy (\%)},
  symbolic x coords={LR,RF,HGB}, xtick=data,
  x tick label style={font=\scriptsize},
  ytick={0,20,40,60,80,100},
  ymajorgrids=true, grid style={gray!25},
  nodes near coords, nodes near coords style={font=\scriptsize},
  enlarge x limits=0.25,
  title={Baseline accuracy},
  title style={font=\footnotesize}
]
\addplot[fill=blue!35,draw=blue!70!black] coordinates
  {(LR,65.2) (RF,68.4) (HGB,70.2)};
\end{axis}
\end{tikzpicture}
\caption{Baseline accuracy values from the PRISM}
\label{fig:baseline}
\end{figure}

Table~\ref{tab:baseline} represents the baseline results available in the
PRISM. Their corresponding run level prediction files are required for
independent verification. Fig.~\ref{fig:baseline} provides a visual
comparison of the 3 baseline models.

\begin{table}[t]
\centering
\caption{Baseline results}
\label{tab:baseline}
\begin{tabular}{lccc}
\toprule
\textbf{Model} & \textbf{Accuracy (\%)} &
\textbf{Brier} & \textbf{ECE}\\
\midrule
Logistic Regression & 65.2 & 0.218 & 0.042\\
Random Forest & 68.4 & 0.198 & 0.058\\
HistGradientBoosting & 70.2 & 0.186 & 0.049\\
\bottomrule
\end{tabular}
\end{table}

\subsection{PRISM Selective Prediction Results}
The PRISM results reports 82.4\% selective accuracy at
88.5\% coverage with Brier score 0.126 and ECE 0.019. The same result
reports a harm prevention value of 92.4\%.

Selective accuracy and full coverage accuracy measure different
quantities. Also the 82.4\% selective accuracy must not be directly
subtracted from or compared with the baseline accuracies in
Table~\ref{tab:baseline} unless the baseline is re-evaluated on exactly the
same accepted subset. A fair comparison should report (i) baseline accuracy
over all test states, (ii) baseline accuracy on PRISM's accepted subset,
(iii) PRISM selective accuracy on that same subset and (iv) coverage.

\begin{table}[t]
\centering
\caption{PRISM metrics}
\label{tab:prism}
\begin{tabular}{lc}
\toprule
\textbf{Metric} & \textbf{Reported value}\\
\midrule
Selective accuracy & 82.4\%\\
Coverage & 88.5\%\\
Brier score & 0.126\\
Calibration error & 0.019\\
Harm prevention & 92.4\%\\
Historical states indexed & 8,665\\
\bottomrule
\end{tabular}
\end{table}

\begin{table}[t]
\centering
\caption{Separate 2025 results~\ref{tab:prism}.}
\label{tab:2025snapshot}
\begin{tabular}{lc}
\toprule
\textbf{Metric} & \textbf{Reported value}\\
\midrule
Baseline accuracy & 80.76\%\\
PRISM selective accuracy & 86.21\%\\
\bottomrule
\end{tabular}
\end{table}

\subsection{Ablation Analysis}

\begin{figure*}[t]
\centering
\includegraphics[width=0.92\textwidth]{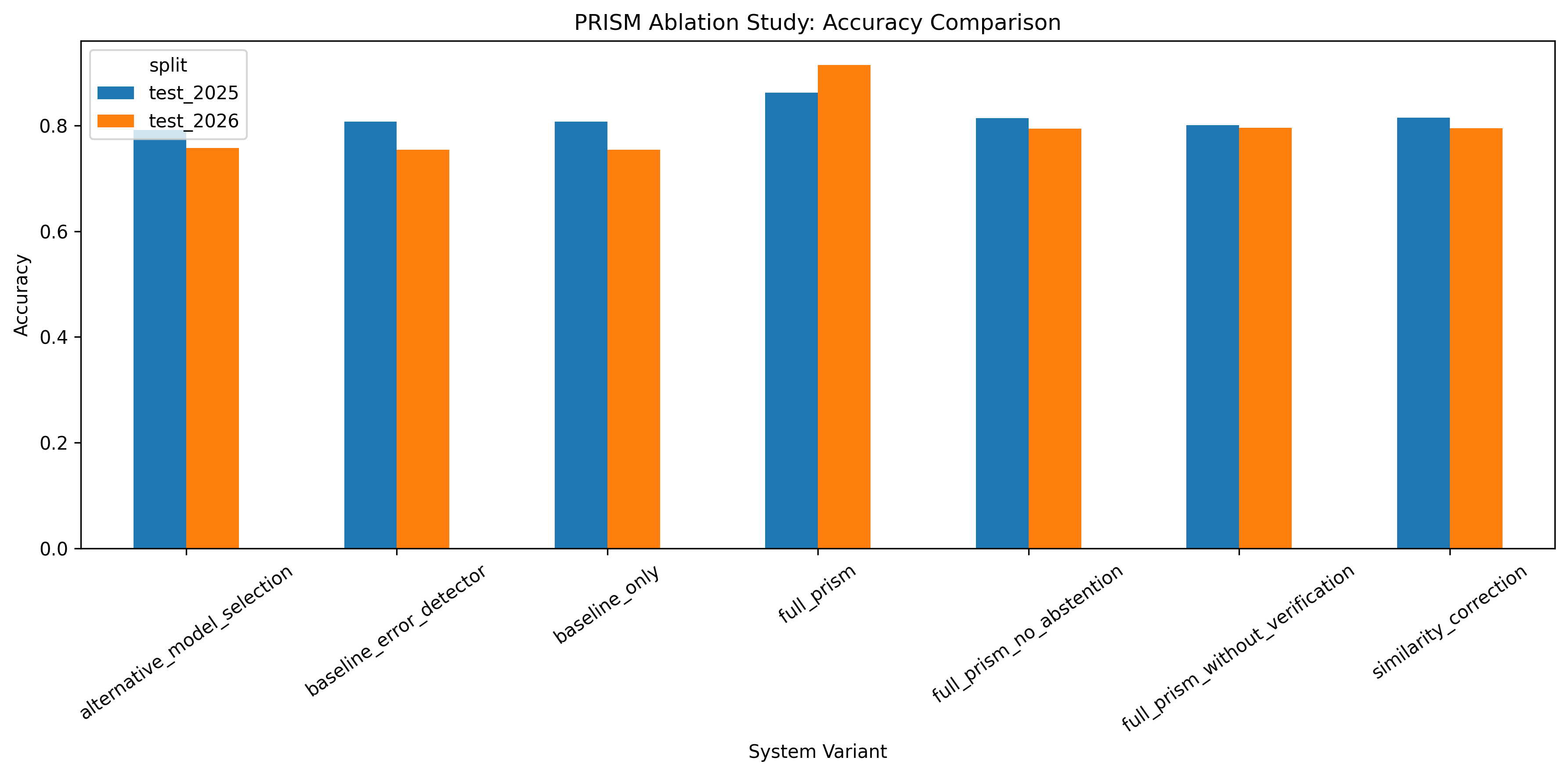}
\caption{Accuracy of PRISM ablation variants on the 2025 and 2026 test splits.}
\label{fig:ablation-accuracy}
\end{figure*}

We conducted ablation experiments to understand how the different PRISM
components contribute to the overall performance. Fig.~\ref{fig:ablation-accuracy}
compares the complete pipeline with its modified variants on the 2025 and
2026 test splits. Full PRISM has the highest accuracy in both cases,
with a more noticeable performance gap in 2026.

\begin{figure}[t]
\centering
\includegraphics[width=\columnwidth]{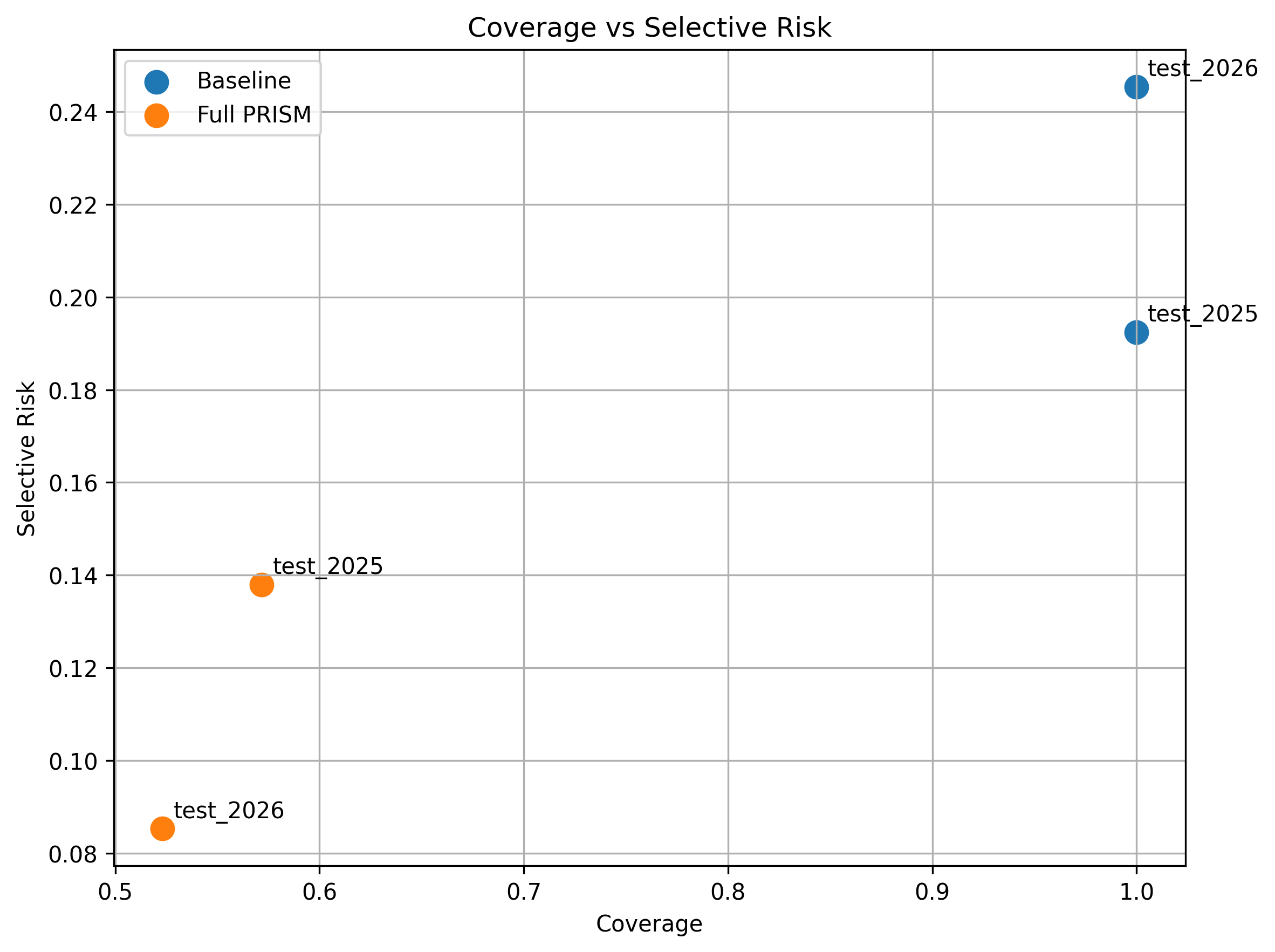}
\caption{Coverage and selective risk of the baseline and full PRISM on the
2025 and 2026 test splits.}
\label{fig:ablation-risk-coverage}
\end{figure}

Accuracy alone does not capture the effect of abstention. Fig.~\ref{fig:ablation-risk-coverage}
compares the coverage and selective risk of the baseline and full PRISM.
The baseline makes predictions for all test examples, while PRISM operates
at lower coverage with lower selective risk in both years. The recorded coverage is approximately 57\% for 2025 and 52\% for 2026.

\section{Discussion and Limitations}

The results are encouraging, but there is still not enough evidence to
conclude that PRISM improves prediction quality across all settings.
Its performance needs to be considered alongside coverage, since higher
selective accuracy may come at the cost of making fewer predictions.
In practice, the usefulness of this trade-off depends on how many
predictions the application can afford to withhold and how reliable the
remaining predictions are. Calibration should also be evaluated separately
for all test samples and for the accepted predictions, as these measure
different aspects of model reliability.

Finally, the experiments focus only on IPL 2nd innings prediction.
Although this provides a useful setting for studying reliability and
self-correction, further evaluation on other tasks and datasets is needed
to determine whether PRISM can be applied more broadly.

\section{Threats to Validity and Reproducibility}
Temporal leakage is a key risk in over level sports prediction. All states from a single match should remain in the same partition and retrieval at test time must exclude future states and any test labels. Similarly, the error risk detector should be trained using out of fold or held out base model predictions. Thresholds should be selected using validation data only and then frozen before evaluating on the out of time seasons.

\section{Conclusion}

This paper presented PRISM, a framework designed to improve the
reliability of model predictions after they are generated. It combines
error risk estimation, diagnosis, retrieval of similar historical cases
and verification of possible corrections. Based on this process, PRISM
can retain the original prediction, replace it with a verified
alternative or abstain when the available evidence is insufficient.

We evaluated the framework using IPL 2nd innings prediction. In the
reported evaluation, PRISM achieved 82.4\% selective accuracy
at 88.5\% coverage with a Brier score of 0.126 and an ECE of 0.019.
These results reflect a particular evaluation and should be
considered alongside the coverage at which they were obtained.

Future work will focus on more consistent evaluation through
matched-subset comparisons, complete ablation reporting and further
robustness testing. Testing PRISM on other datasets and prediction
tasks will also help establish whether the approach generalises beyond
IPL match prediction.

\balance

\end{document}